# A new strategy for achieving multiple continuous cooling stages in an adiabatic demagnetization refrigerator

Yanan Li[1,2], Ke Li[1], Teng Pan[1,2], Yaxuan Wang[1,2], Wei Dai[1,2*]

[1] State Key Laboratory of Cryogenic Science and Technology, Technical Institute of Physics and Chemistry, Chinese Academy of Sciences, Beijing 100190, China

[2] University of Chinese Academy of Sciences, Beijing 100049, China

**Abstract**

This paper presents a three-stage continuous adiabatic demagnetization refrigerator (CADR) simultaneously providing cooling platforms at two different temperatures. Unlike conventional CADR with two continuous stages, this system does not require an extra continuous stage. It is achieved with carefully designed heat switch operations and stage sequences. This new strategy significantly reduces system complexity and mass. The detailed implementation of the two cooling platforms is described, and factors limiting system performance are analyzed. This CADR system has achieved a cooling power of 20 μW@1 K and 4 μW@300 mK simultaneously, offering a new design strategy for the development of CADR with multiple continuous stages.

## 1. Introduction

Space astronomy missions and specific ground applications have prompted the rapid development of adiabatic demagnetization refrigerators (ADR). They are especially favored for space missions due to gravity-insensitivity, high reliability, and precise temperature control. Inside an ADR, a salt pill is magnetized to reject heat to a heat sink through a heat switch (the regeneration process), followed by demagnetization to absorb the heat from the load (the cooling process). A conventional ADR generally follows an reverse Carnot cycle and operates in a non-continuous way, which seriously limits its specific cooling power density. The advent of continuous ADR (CADR), especially the configuration proposed and realized by Shirron et al[1], greatly improves the situation and can lead to a more compact system given the same cooling power requirement. With the ever-increasing demand for scientific explorations, CADR with extra continuous cooling stages at some middle temperatures has further been developed to meet the different operational temperatures required for different instruments or optics in a single system.

Fig. 1 shows two present different four-stage CADR configurations, which can provide two cooling platforms at different temperatures $T_{C1}$ and $T_{C2}$. Inside the configurations, two typical strategies for realizing continuous cooling are illustrated, i.e., the series strategy and the parallel strategy. For the series strategy, e.g., stage 3 and stage 4 in both Fig.1a and Fig.1b, the stage 4 keeps a constant temperature during both demagnetization and magnetization processes. During the demagnetization process, the heat switch in between them is at OFF state and the cooling power is provided by stage 4. As its cooling capacity gets depleted, the heat switch turns to ON-state and stage 4 starts to magnetize (the regeneration process). Meanwhile, stage 3 begins to demagnetize to generate the cooling power to both absorb the heat rejected by stage 4 and cools the load. In the way, a continuous cooling power can be provided at $T_{C2}$. For the parallel configuration, stage 1 and 2 in Fig.1b, it is more straightforward. These two identical stages operate alternatingly to provide continuous cooling power at $T_{C1}$.

The above example illustrates how multiple continuous cooling stages can be provided. In practice, a five-stage CADR is developed to provide continuous cooling at 1 K and 50 mK by NASA for constellation-X mission[2]. CEA has also proposed a 5-stage CADR design for Athena mission, targeting continuous cooling at 1.8K and 325mK, as well as operation at 50mK with an 80% duty cycle[3]. Both these two designs adopt the series configuration for the two continuous stages. JAXA, NASA, and CEA have collaborated to design a seven-stage CADR aimed at providing continuous cooling at 1.75 K, 300 mK, and 100 mK for the LiteBIRD mission[4]. It employs the parallel configuration for the 1.75 K continuous stage, while the other continuous stages adopt the series configuration. The above designs always require the introduction of an additional ADR stage to provide an extra continuous cooling platform at a middle temperature. While each ADR stage consists of a superconducting magnet, heavy magnetic shield as well as a demanding power source, influence due to the increased complexity and overall mass needs to be carefully considered in the design stage.

In reference to the systems illustrated in Fig.1, here we propose and have experimentally verified a different strategy to realize the same function with only three stages, which leads to a much simplified configuration.

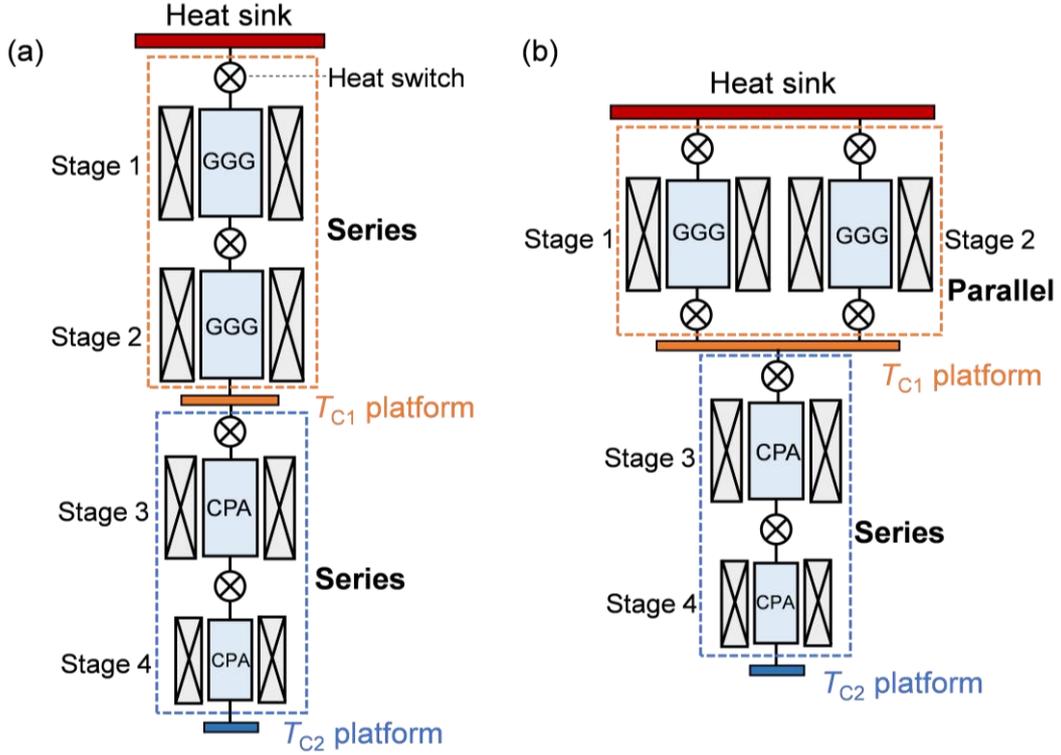

Figure 1. Schematic of typical CADR providing two continuous cooling at two different temperatures: (a) series strategy for the two continuous stages; (b) parallel strategy for continuous stage at $T_{C1}$ and series configuration for continuous stage at $T_{C2}$.

## 2. System overview

The schematic of the novel strategy[5] and photo of our three-stage CADR setup based on this strategy are shown in Fig.2. The continuous stage 3 is connected to stage 2 via HS4 on the one side, while the other side is directly connected to the $T_{C2}$ platform, here 300 mK. This is a typical series strategy as explained above. However, continuous cooling at $T_{C1}$, here 1K, is provided through a combined operation of stage 1 and stage 2 as well as coordinated operating of heat switches HS2 and HS3. Typically, this configuration can provide the cooling power at 1 K and 300 mK, simultaneously. Before delving into the details of the strategy, we first introduce the hardware configuration and experiment instrumentation. Then with typical cooling curves, we will explain how the strategy works.

A GM-type pulse tube cryocooler provides a 3.5 K heat sink for the CADR, and all three superconducting magnets together with their magnetic shields are thermally anchored to this platform. Two different refrigerant materials are used in our CADR: gadolinium gallium garnet (GGG) for stage 1, chromic potassium alum (CPA) for stage 2 and 3. All salt pills are supported by 3D-printed PEEK suspensions[6]. Some key design and operating parameters related to the CADR are listed in Table 1.

All heat switches are active gas-gap heat switches (HS), which are well suited to our operating temperatures and allow easy switching between ON and OFF states[7]. However, it is also due to this reason, the lowest cooling temperature that can be provided is limited to be above 200 mK, at which the gas pressure inside is too low to ensure a useful thermal conductance. To realize a much lower temperature needs the use of superconducting heat switch or magneto-resistance heat switch.

Ruthenium oxide resistors (Rox[TM]) are installed on each stage and cooling platforms for temperature monitoring. Cernox[TM] resistance cryogenic temperature sensor is installed on the 3.5 K platform. The temperatures are measured using the Lake Shore Model 372 and 224 AC resistance bridges. Two heaters with a resistance of 1kΩ, powered by TH6402B DC power sources, are respectively installed on the two cooling platforms to simulate the heat load. Lake Shore Model 625 magnet power supplies are used to drive the superconducting magnets of each stage. A LabVIEW program communicates with the instrumentations to acquire temperature and magnet current data, etc, and employs PID control routines to adjust the current ramp rate to maintain the target temperature.

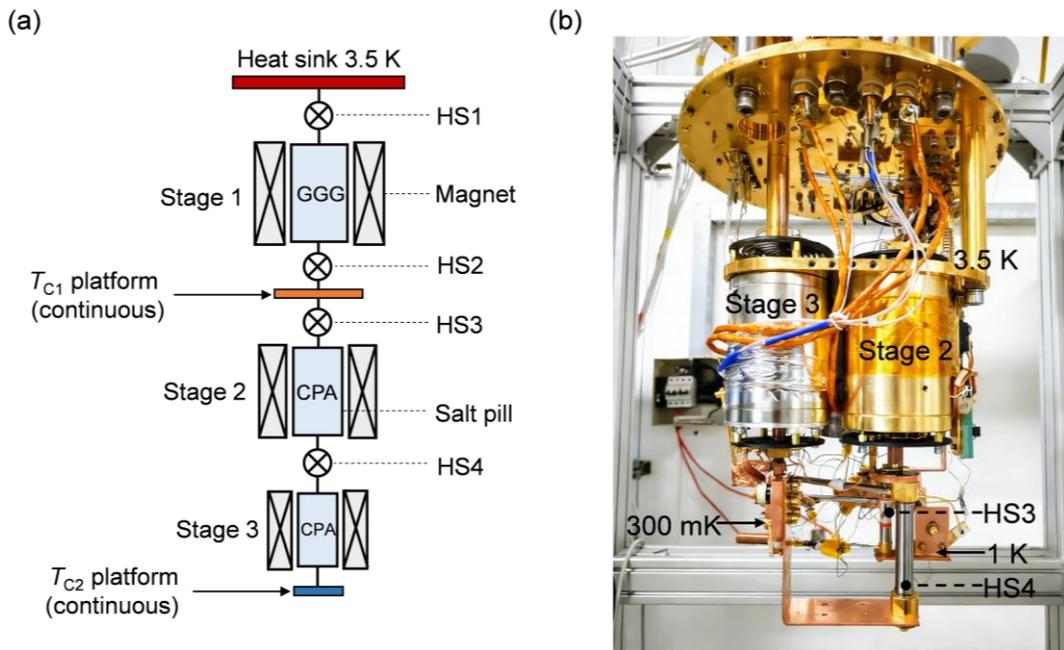

Figure 2. (a) Schematic of our new strategy. (b) Photo of three-stage CADR providing continuous cooling at 1 K and 300 mK. Some visible components have already been labeled in the photo.

Table 1. CADR design and operating parameters

| Stage | Refrigerant (g) | Magnet field (T) | Max current (A) | Upper $T$ (K) | Lower $T$ (K) |
|---|---|---|---|---|---|
| 1 | GGG: 300 | 4 | 10 | 4.5 | 1 |
| 2 | CPA: 98 | 2.3 | 17.7 | 1.2 | 0.23 |
| 3 | CPA: 45 | 0.3 | 1.6 | 0.3 | 0.3 |

## 3. Experimental results and discussions

Warm Start

Fig.3 demonstrates typical temperature and magnetic field profiles over the full period. A small fraction at the leftmost part shows the warm start process (from $t_a$ to $t_b$). Stage 3 (blue line) starts from zero magnetic field and is cooled sequentially by the demagnetization of stage 1 and stage 2, which ends at around 1 K and 230 mK, respectively. After approximately 50 minutes, the system transitions into the continuous cooling process. In correspondence to these curves, the system has achieved a cooling power of 20 µW@1 K and 4 µW@300 mK simultaneously. The following introduces in detail the cooling processes for 1 K and 300 mK platforms, respectively. For clarity, we sometimes neglect the description of transitional processes of the heat switches (normally takes ~ minutes to over 10 minutes), though they are illustrated in the figures.

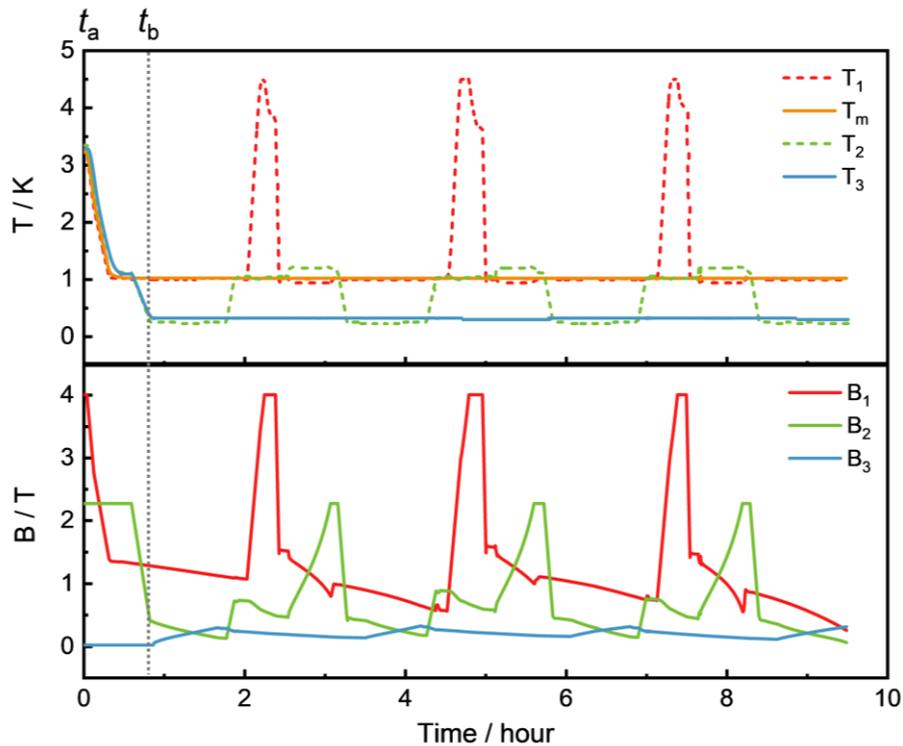

Figure 3. Temperature and magnetic field profiles over the full period with the cooling power of 20 µW@1 K and 4 µW@300 mK. The period from $t_a$ to $t_b$ refers to the warm start process. $T_1$, $T_m$, $T_2$, $T_3$ refer to the temperature of stage 1, 1 K platform, stage 2 and stage 3 (300 mK platform); $B_1$, $B_2$, $B_3$ refer to the magnetic field of stage 1, stage 2 and stage 3.

### Continuous cooling at 300 mK

It is accomplished by stage 2 and stage 3 with series strategy, as explained in introduction section. A breakdown view as well as the typical curves from Fig.2 are show in Fig.4. Stage 3 is directly connected to the 300 mK platform and is either demagnetized or magnetized as needed to maintain a steady 300 mK platform. The cooling power for regenerating stage 3 is supplied by stage 2, which is held at a lower temperature of 230 mK and simultaneously absorbs the heat load on the 300 mK platform. During this phase, HS3 is in the ON state. Once stage 2 is nearly depleted, HS3 is turned off and stage 2 is warmed up to deposit the heat.

The cooling temperature of stage 2 is chosen to be at 230 mK is due to the following consideration. Once the cooling temperature of stage 2 is fixed, the heat transfer in between stage 2 and stage 3 is proportional to the product of on-state conductance of HS4 and the temperature difference[8]. The lower this temperature, the lower on-state conductance with a bigger temperature difference. This value is strongly related to the HS4 performance and 230 mK is thus experimentally determined to maximize heat transfer rate. The on-state conductance of HS4 at this temperature is only approximately 0.6 mW/K. From Fig.3(b), it can also be observed that a significant time lag exists between the initiation of stage 2 demagnetization (point a) and stage 3 magnetization (point b). This means that the transition of HS4 from OFF to ON takes as long as 13 minutes, which increases the cycle period and limits the cooling power of the 300 mK platform. This phenomenon is abnormal and certainly needs improvement.

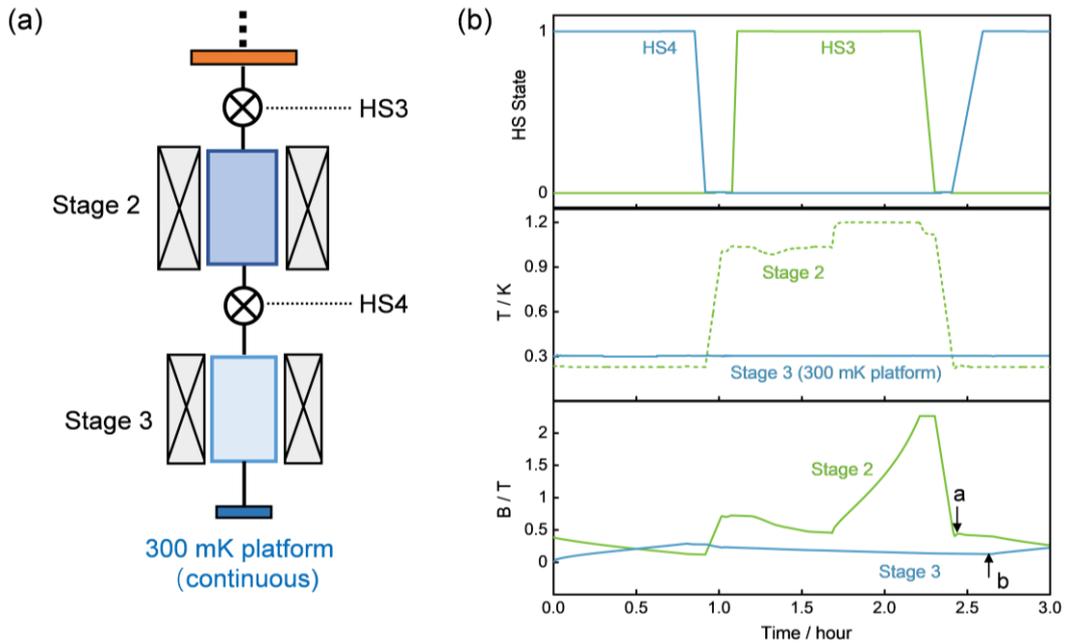

Figure 4. (a) Schematic of the lower two stages in our CADR; (b) Heat switch state, temperature and magnetic field during one cycle for 300 mK continuous cooling. For HS state, 1 indicates ON and 0 indicates OFF, and the slopes represent the switching of the heat switches.

Continuous cooling at 1 K

As said before, it is realized by stage 1 and 2 with the appropriate regulation of HS2 and HS3. A breakdown view as well as the typical curves from Fig.2 are shown in Fig.5. For clarity, it only shows the states of the two key heat switches, HS2 and HS3. Table 2 further gives the states of three related heat switches.

The duty of continuous cooling at 1 K is fulfilled sequentially by two parts : stage 2 duty (from $t_a$ to $t_b$) and stage 1 duty (the rest). The following first introduce the stage 2 duty. When cooling capacity of stage 2 is depleted at 230 mK after providing the cooling at 300 mK, as mentioned above, it is magnetized and warms up to around 1 K under a relatively high magnetic field at $t_a$. Subsequentially, HS3 is turned fully ON and stage 2 takes over the cooling duty of the 1 K platform. During stage 2 duty, stage 1 must complete its regeneration to prepare for its own upcoming duty. The specific procedure is as follows: once HS3 is fully conducting, HS2 is turned off, and stage 1 is magnetized and warms up to a temperature above the heat sink (about 4.5 K) to undergo a rapid regeneration, releasing its magnetization heat to the heat sink via HS1.

Following stage 2 cooling duty of the 1 K platform comes the stage 1 cooling duty. HS1 is turned off and stage 1 is demagnetized again to cool down near 1 K at $t_b$. HS2 is turned on, and stage 1 resumes its cooling duty. Meanwhile, HS3 remains ON state and stage 2 warms up to 1.2 K, slightly higher than the 1 K platform, to complete its regeneration. During the stage 1 duty (from $t_b$ to $t_c$), it provides cooling to the 1 K platform and simultaneously absorbs the magnetization heat released by stage 2. Upon stage 2 completing its regeneration at $t_c$, HS3 is turned OFF. Stage 2 is demagnetized to 230 mK again to serve for the 300 mK continuous cooling operation. And stage 1 continues its cooling duty until the cooling capacity of stage 2 is depleted at 230 mK. With this, one continuous cooling cycle for the 1 K platform is finished.

It is important to note that, in order to maintain temperature stability of the 1 K platform, during the transition of HS2 and HS3, one heat switch must be completely turned ON or OFF before the switching of the other can begin.

Through experiments, the on-state conductance of HS1, HS2, and HS3 are estimated to be 35 mW/K@3.5 K, 10 mW/K@1 K, and 4.6 mW/K@1 K, respectively. Compared to HS2, HS3 is an old heat switch in our laboratory, and its lower on-state conductance limits both the regeneration rate of stage 2 and the cooling power of the 1 K platform.

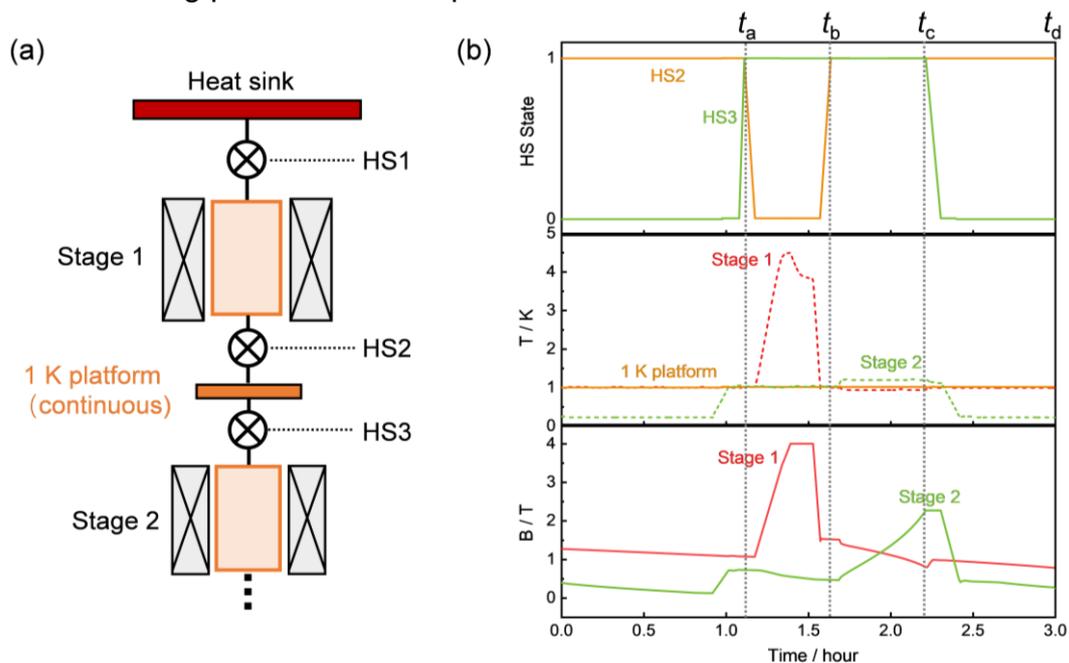

Figure 5. (a) Schematic of the upper two-stage CADR; (b) Heat switch state, temperature and magnetic field during one cycle for 1 K continuous cooling. For HS state, 1 indicates

ON and 0 indicates OFF, and the slopes represent the switching of the heat switches. From $t_a$ to $t_b$ is stage 2 cooling duty, and the rest is stage 1 cooling duty.

Table 2. States of heat switches during the continuous cooling process at 1 K.
(The switching transients are ignored, and only the primary states are listed.)

| Heat switch | Stage 2 duty | Stage 1 duty | | |
|---|---|---|---|---|
| | $t_a$-$t_b$ | $t_b$-$t_c$ | $t_c$-$t_d$ | $t_d$-$t_a$ |
| HS1 | ON | OFF | OFF | OFF |
| HS2 | OFF | ON | ON | ON |
| HS3 | ON | ON | OFF | OFF |

## 4. Conclusion

In summary, a new strategy for achieving multiple continuous stages in an CADR has been proposed and experimentally demonstrated. A three-stage CADR achieves a cooling power of 20 µW@1 K and 4 µW@300 mK simultaneously. The cooling powers can be adjusted as needed with different operating parameters of each stage. Compared with a conventional CADR providing two continuous cooling stages, this design shows an apparently reduced complexity. Presently, the parameters of each stage have not been optimized and the CADR operation is just manually regulated to demonstrate the concept. Improvements of the heat switches are also needed to further enhance the cooling performance.

## Acknowledgments

This work is financially supported by the National Key Research and Development Program of China (Grant No. 2021YFC2203303).

## Author Statement

**Yanan Li:** Investigation, Methodology, Validation, Visualization, Writing - original draft, Writing- review & editing. **Ke Li:** Investigation, Software, Supervision. **Teng Pan:** Resources, Investigation. **Yaxuan Wang:** Investigation. **Wei Dai:** Conceptualization, Methodology, Project administration, Writing- review & editing

## References

[1] Shirron PJ, Canavan ER, DiPirro MJ, Tuttle JG, Yeager CJ. A Multi-Stage Continuous-Duty Adiabatic Demagnetization Refrigerator. Advances in Cryogenic Engineering, 2000: 1629–38.
[2] Shirron P, Canavan E, DiPirro M, Francis J, Jackson M, Tuttle J, et al. Development of a cryogen-free continuous ADR for the constellation-X mission. Cryogenics, 2004, 44: 581–8.
[3] Duval JM, Bancel F, Charles I, Durand JL, Léon A, Lizion J, et al. 5-Stage ADR Cooler for the Athena Space Mission: Design and Preliminary

Characterization. International cryocooler Conference, Madison, Wisconsin, June 3-6, 2025.

[4] Duval J-M, Prouvé T, Shirron P, Shinozaki K, Sekimoto Y, Hasebe T, et al. LiteBIRD Cryogenic Chain: 100 mK Cooling with Mechanical Coolers and ADRs. Journal of Low Temperature Physics, 2020, 199: 730–6.

[5] Dai W, Liu P, Li K, Li Y. An adiabatic demagnetization refrigerator with multiple continuous cooling stages. CN117168015A, 2023.12.05.

[6] Li Y, Liu P, Zhao P, Li K, Dai W. 3D printed suspension and its evaluation in an ADR refrigerator. Cryogenics, 2024, 140: 103859.

[7] DiPirro MJ, Shirron PJ. Heat switches for ADRs. Cryogenics, 2014, 62: 172–6.

[8] Shirron PJ. Passive gas-gap heat switches for use in adiabatic demagnetization refrigerators. AIP Conference Proceedings, AIP Conference Proceedings, 2002, 613: 1175-82.